\documentclass[12pt]{article}
\usepackage{setspace}
\usepackage{fullpage}
\usepackage[english]{babel}
\usepackage[T1]{fontenc}
\usepackage{graphicx}
\usepackage{amssymb,amsthm,amsmath,,amsfonts}
\usepackage{wasysym}
\usepackage{ae}   

\usepackage{aecompl} 
\usepackage{float}
\usepackage{natbib}
\usepackage[font=small,labelsep=none]{caption}

\renewcommand{\deg}{^\circ}

\newcommand{\arcsec}{^{\prime\prime}}

\newcommand{\DSH}{$D_{SH}$}
\newcommand{\Dc}{$D_{cluster}$}
\newcommand{\Dp}{$D_{pair}$}
\newcommand{\ie}{{\it i.e.}}
\newcommand{\eg}{{\it e.g.}}
\newcommand{\etc}{{\it etc.}}

\begin{document}

%\doublespacing
%\citationstyle{agsm}
%\citationmode{abbr}

\mbox{ }
\vspace{2cm}
\begin{center}
{\bf{\Large Searching for the first Near-Earth Object family}} \\

\vspace{0.5cm}
Eva Schunov\'a${}^\mathrm{a}$, Mikael Granvik${}^\mathrm{b,e}$, Robert Jedicke${}^\mathrm{b}$, Giovanni Gronchi${}^\mathrm{c}$, Richard Wainscoat${}^\mathrm{b}$, Shinsuke Abe${}^\mathrm{d}$\\

\vspace{0.5cm}

E-mail: schunova@ifa.hawaii.edu \\

\vspace{0.5cm}

${}^\mathrm{a}$Department of Astronomy, Physics of the Earth and Meteorology, Comenius
University, Mlynsk\'a dolina, Bratislava, 942 48, Slovakia \\
${}^\mathrm{b}$Institute for Astronomy, University of Hawaii, 2680 Woodlawn Drive,
Honolulu, HI 96822 \\
${}^\mathrm{c}$Dipartimento di Matematica, University of Pisa, Piazzale B. Pontecorvo 5, 56127 Pisa, Italy \\
${}^\mathrm{d}$Institute of Astronomy, National Central University, No. 300, Jhongda
Rd., Jhongli, Taoyuan, 32001, Taiwan. \\
${}^\mathrm{e}$Department of Physics, P.O. BOX 64, 00014 University of Helsinki, Finland \\

\vspace{0.5cm}

%Submitted to \emph{Icarus}\\
Submitted \today\\
%Revised \\
\vspace{1cm}
Manuscript pages: 36 \\
Tables:  3 \\
Figures: 9 \\

\vspace{1cm}

\end{center}

\newpage

\mbox{ }
\vspace{0cm}
\begin{center}
{\bf Proposed Running Head:} Searching for the first NEO Family \\
\end{center}
\vspace{10cm}
{\bf Editorial correspondence to:} \\
Eva Schunova \\
Institute for Astronomy \\
University of Hawaii \\
2680 Woodlawn Drive \\
Honolulu, HI 96822 \\
Phone: +1 808 294 6299 \\
Fax: +1 808 988 2790 \\
E-mail: schunova@ifa.hawaii.edu 

\newpage

\begin{abstract}

We report on our search for genetically related asteroids amongst the
near-Earth object (NEO) population --- families of NEOs akin to the
well known main belt asteroid families.  We used the technique
proposed by \citet{fu2005} supplemented with a detailed analysis of
the statistical significance of the detected clusters.  Their
significance was assessed by comparison to identical searches
performed on 1,000 `fuzzy-real' NEO orbit distribution models that we
developed for this purpose. The family-free `fuzzy-real' NEO models
maintain both the micro and macro distribution of 5 orbital elements
(ignoring the mean anomaly). Three clusters were identified that
contain four or more NEOs but none of them are statistically
significant at $\ge3\sigma$.  The most statistically significant
cluster at the $\sim2\sigma$ level contains 4 objects with $H<20$ and
all members have long observational arcs and concomitant good orbital
elements.  Despite the low statistical significance we performed
several other tests on the cluster to determine if it is likely a
genetic family.  The tests included examining the cluster's taxonomy,
size-frequency distribution, consistency with a family-forming event
during tidal disruption in a close approach to Mars, and whether it
is detectable in a proper element cluster search.  None of these
tests exclude the possibility that the cluster is a family but neither
do they confirm the hypothesis.  We conclude that we have not
identified any NEO families.

\end{abstract}

{\bf Key Words:} NEAR-EARTH OBJECTS, ASTEROIDS, DYNAMICS
\newpage

\section{Introduction}
\label{sec:introduction}

We report here on our null search for a statistically significant cluster
of genetically related near-Earth objects (NEO).  The identification
of such a NEO `family' would enable further research into the physical
characteristics of NEOs as was the case with the identification of
asteroid families in the main belt, Jupiter Trojan and Trans-Neptunian
minor planet populations.  The physical characteristics of the
ensemble of NEO family members --- taxonomic and
mineralogical types, sizes, rotation periods, shapes, pole
orientations, existence of satellites, \etc --- would lead to a better
understanding of their morphology and the mechanisms affecting their dynamical and
collisional evolution.  If the family members have a small minimum
orbital intersection distance (MOID) with the Earth (or any other
planet) their presence on a sub-set of dangerous orbits will increase
the total impact probability over what is currently understood because
the enhancement in orbit element phase space is not incorporated into
contemporary NEO models.

In one sense there are already some known NEO families since
associations have been proposed between several meteor showers and
their presumed parent body NEOs \citep[both asteroids and comets,
  \eg][]{sek1973,por2004}.  The most known and well-accepted is the connection between the unusual B-type NEO (3200)
Phaethon and the Geminid meteor complex \cite{oht2008}. However, the associations between meteors and NEOs are not
families in the traditional undestanding.
The known asteroid families are produced through the catastrophic
disruption of a parent asteroid in a severe impact with another
asteroid.  The discovery of genetically related pairs of asteroids in
the main belt \citep[\eg][]{vok2008} suggests that other mechanisms
may produce asteroid families such as the spin-up and rotational
fission of rapidly rotating objects or the splitting of unstable
binary asteroids.  Meteoroids are the small-end tail of those family
creation mechanisms but can also be produced through surface ejection
driven by the volatilization of sub-surface ices on comets or `active
asteroids'.  
While no formal distinction exists between the meteor
`families' and the asteroid families the size ratio between the
largest and second largest fragments nicely divides the samples and
creation mechanisms.

No NEO families have been identified and confirmed.  \citet{dru2000} suggested that there were many `associations' in the NEO population but \citet{fu2005} showed that all of them were likely chance alignments of their orbits.  The dearth of NEO families is in contrast to the more than 50 families known in the main belt \citep[\eg][]{nes2005}.  Indeed, \citet{hir1918} proposed the existence of main belt asteroid families when there were only 790 known main belt asteroids while there are now more than 7,500 known NEOs. 

Asteroid families are typically identified by the similarity of the family member's orbital elements.  Several metrics have been developed to quantify the orbital similarity \cite[\eg][]{sou1963,val1999,jop2008,vok2008}.  The best known is the Southworth-Hawkins D-criterion \citep[\DSH;][]{sou1963} that was developed and used to search for parent comets of meteor streams and to link meteor streams with NEOs \citep[\eg][]{gaj2005,oht2008}.   

It is important to keep in mind that simply identifying members within a population with similar orbital elements does not mean that they are actually a genetically related family.  \ie\ that they do not necessarily derive from a single parent object.  As the number of objects in the population increases there is a corresponding increase in the likelihood that chance associations will mimic families and extra care must be taken to establish a proposed family's statistical significance.  

The major difference between the ability to identify families in the
NEO and main belt populations is that the orbits of asteroids in the
main belt are stable on timescales comparable to the age of solar
system.  Their long term stability allows the calculation of most
objects' proper elements \cite[\eg][]{kne2002}, essentially
time-averaged orbital elements, that are better suited to the
identification of similar orbits that are the hallmarks of asteroid
families.  Furthermore, the non-gravitational evolution of the main
belt asteroids' orbits is slow under the influence of the Yarkovsky
effect \citep[\eg][]{vok2006a} so that they occupy the same orbital
element phase space for long periods of time.

The main belt families identified by the similarity in their orbit
elements range in age from millions to billions of years.
Spectroscopic campaigns have confirmed that the members of older
families typically share the same spectral type; as expected if they
all derive from the same parent body and/or are covered by similar
regolith during the family formation event
\citep[\eg][]{cel2002,izv2002a}.  
Near Earth objects on the other hand reside in a turbulent dynamical
environment with average lifetimes of $\sim 10^6$~yr under the strong
gravitational influence of the terrestrial planets and Jupiter
\citep[\eg][]{mor2002,gla1997}.  Thus, the calculation of NEO proper
elements is difficult because they may cross the orbits of the
planets during their time evolution \citep{gro2001}.  The resulting NEO proper elements are only valid for the
time between their close encounters with the planets and when mean
motion resonances of low order with the planets do not occur.

The search for similar orbits amongst the NEO population
using only the osculating elements is even more limited in the time
frame over which the orbits maintain their coherence \citep{pau2005}.
Furthermore, the non-gravitational accelerations acting on NEOs are generally much
stronger than those acting on the more distant asteroids of the solar
system.  Thus, if a NEO family exists it will only be possible to
identify it for a short time period after its creation as its members
rapidly disperse.

NEO families might form by several imaginable methods \citep{fu2005}:
\begin{itemize}
\item spontaneous disruption (\eg\ YORP spin-up, internal thermal stresses)
\item tidal disruption during close approach to a planet
%\item family-producing collisions between main belt objects near a main belt resonance followed by rapid dynamical evolution of some of the members onto NEO orbits
\item intra-NEO collisions
\item collision with a smaller main belt asteroid
\end{itemize}

It is difficult or impossible to assign a likelihood to the different formation mechanisms.  While 11\% of comets spontaneously disrupt \citep[\eg][]{mas2009} as they enter the inner solar system $(1.0 AU < q \leq 1.5 AU)$, presumably due to thermal stresses induced by approaching the Sun, the NEOs have typically resided in the terrestrial planet zone for long periods of time and many orbital periods.  They are thus unlikely to disrupt by this mechanism.  On the other hand, simulations suggest that the YORP thermal torque can increase an asteroid's spin rate to the point where it spontaneously sheds material and some of this material may be ejected faster than the parent body's escape speed \citep[\eg][]{wal2006,pra2010,jac2011}.  If the process is repeated multiple times over a short time span it might be possible to create an asteroid family in this manner.  Families created in this way might be distinguished by their rotation rates, mass ratios, or pole orientations of their members.

Tidal disruption of NEOs may occur when they pass close to a planet \citep[\eg][]{rich1998,wal2006,wal2008} as occurred in the production of the family of objects associated with Comet Shoemaker-Levy-9 \citep[][]{sek1994}.  While the tidal disruption of asteroids has been simulated under a variety of conditions (\eg, spin rate, pole orientation, closest approach distance, encounter speed), there is no estimate of how many tidal disruptions actually occur.

The likelihood of forming NEO families with the third mechanism is probably relatively small, since the space number-density of NEOs is very small compared to the space number-density of main belt asteroids \citep{bot1996}.

Perhaps the most likely formation mechanism for a NEO family is a collision between the NEO and a smaller main belt asteroid \citep{bot1996}.  Many NEOs remain on orbits with aphelia in or beyond the main belt where they are typically slower than other asteroids at the same distance and therefore suffer higher speed collisions compared to those between two main belt asteroids at the same heliocentric distance.  Thus, smaller projectiles could be effective impactors at catastrophically disrupting a NEO.  

In summary, NEO families will provide to the opportunity of exploring the physics of asteroid disruptions at different scales and for different reasons than those observed in the main belt.

\section{Method}
\label{sec:method}

We searched for families in the NEO population using the method
proposed by \citet{fu2005}.  The technique identifies subsets of objects with similar orbits within a population.
We used their osculating elements, because of the problems involved in calculating NEO proper elements.  We are
thus limited to identifying NEO families that have formed relatively
recently though this is not really a problem because 1) NEO dynamical
lifetimes\footnote{A NEO's dynamical lifetime is the time during which
  it remains in the NEO region.} are on the order of a million years
and 2) young families are interesting.

We used the Southworth-Hawkins \DSH\ criteria (\S\ref{sec:dcriterion}) to quantify the similarity between two orbits.  It could be argued that other criteria would yield better results but this criteria has been well tested and forms the basis of the \citet{fu2005} technique.  A set of NEOs that all have mutual \DSH\ criteria below a threshold \Dc\ is considered a `cluster'.  The technique allows for and favors sub-clustering within the cluster under the assumption that there could be a tight `core' within a cluster surrounded by a looser assemblage of related objects.  
We used the Minor Planet Center (MPC) mpcorb.dat orbit element data set that contained 7,563 NEOs as of March 2011 (through 2011~DW ). The technique is described in detail below.

The real difficulty in identifying NEO families is not the identification of the clusters but in establishing their statistical significance.  Indeed, as shown by \citet{fu2005} for NEOs and \citet{pau2005} for fireballs, it is surprisingly difficult to prove that similar orbits are statistically significant within a population.  We first tested our method on synthetic family-free NEO orbit models but then developed more realistic family-free NEO models derived from the known NEO population.  We used multiple instances of the family-free NEO models to establish the statistical significance of our NEO clusters.

\subsection{The Southworth-Hawkins $D$-criterion}
\label{sec:dcriterion}

The Southworth-Hawkins \DSH\ metric quantifies the similarity of two orbits using five orbital elements \citep{sou1963}.  We use the \DSH\ metric in its original form because it has an established pedigree (\eg\ searching for parent comets of meteor streams) that has been used to successfully identify
members of known meteor streams \citep{sou1963,sek1970} and main belt asteroid
families \citep{lin1971}.
 
The \DSH\ metric between two orbits denoted with subscripts $m$ and $n$ is defined by:
\begin{equation}
 D_{SH} = \sqrt{{d_1}^2 + {d_2}^2 + {d_3}^2 + {d_4}^2} ,
 \label{eq:DSH}
\end{equation}
with
\begin{eqnarray}
 d_1 &=& \frac{q_m-q_n}{AU}, \\
 d_2 &=& e_m-e_n,  \\
 d_3 &=& 2\sin{(I/2)},  \\
 d_4 &=& (e_m+e_n)\sin{(\Pi/2)}
\end{eqnarray}
\begin{eqnarray}
I &=& \arccos{\left[\cos{i_m}\cos{i_n}+\sin{i_m}\sin{i_n}\cos{(\Omega_m-\Omega_n)}
\right]}, \\
\Pi &=&
(\omega_m-\omega_n)\pm2\arcsin{\left[\cos{\frac{i_m+i_n}{2}}\sin{\frac{
\Omega_m-\Omega_n}{2}}\sec{\frac{I}{2}}\right]}, \label{eq:Pi}
\end{eqnarray}\
\noindent where $q=a(1-e)$ is the perihelion distance, \emph{a} is the
semi-major axis, \emph{e} is the eccentricity, \emph{i} the inclination, $\Omega$
is the longitude of the ascending node and $\omega$ is the argument of perihelion.  \emph{I} represents the angle between the poles of the two orbits and $\Pi$
represents the angle between their perihelia directions \citep{dru2000}.  Use the positive sign for the
$\arcsin$ term in eq.~\ref{eq:Pi} when $|\Omega_m-\Omega_n| \le 180\deg$ and the negative sign otherwise.

\subsection{Identifying clusters in orbital element space}
\label{sec:clustersearch}

We adopted the cluster identification method developed by \citet{fu2005} that in turn incorporated the techniques described by
\citet{dru2000}.  A `cluster' is a grouping of objects with mutually similar orbits whereas we will reserve the term `family' for a cluster with high statistical significance that contains members that are likely genetically related.  \citet{fu2005} showed in limited testing that the method is capable of identifying synthetic NEO families with minimal contamination. 

We grouped objects into clusters based on the values of 4 parameters that are described in greater detail below:
\begin{itemize}
\item \Dc: The maximum \DSH\ between any pair of objects in a cluster. 
\item \Dp: The maximum \DSH\ between `tight' pairs in a cluster. 
\item $SCR_{min}$: The minimum String length to Cluster size Ratio.  
\item $PF_{min}$: The minimum fraction of pairs of asteroids in the cluster with $D_{SH}<D_{pair}$.
\end{itemize}

We identified candidate clusters as sets of $N$ objects with mutual $D_{SH}<D_{cluster}$.  Within each candidate cluster we then identified all pairs ($n$) of asteroids with $D_{SH}<D_{pair}$. The pair fraction $PF$ is the number of detected
pairs divided by the number of all possible pairs in the cluster
($PF=n/\frac{N(N-1)}{2}$) and we required that all our clusters satisfy $PF>PF_{min}$.  Then, within each candidate cluster we determined the maximum string length $L$ --- the maximum number of objects that are connected in a continuous pair-wise fashion such that each sequential pair in the string satisfies $D_{SH}<D_{pair}$.  The $SCR$ is the ratio of the number of objects in the string to the number of objects in the cluster, $SCR=L/N$, and our final set of clusters all satisfy $SCR>SCR_{min}$.  Note that a string must contain more than two objects.

The goal of our selection criteria is to identify tight clusters of objects in orbital element space but the $PF$ and $SCR$ cuts recognize that as NEOs undergo rapid dynamical and non-gravitational evolution some of the members may evolve quickly onto different orbits.  The pair and string searches allow for a tight `core' of objects with a periphery of other objects.

\subsection{Selecting thresholds for the cluster identification algorithm}
\label{sec:SelectingThresholds}

While it is simple to verify that our algorithm can identify objects in clusters with similar orbital elements it is not trivial to select the four threshold parameters (\Dc, \Dp, $SCR_{min}$, $PF_{min}$) to maximize the cluster detection efficiency while minimizing the contamination by false positives.  
This required a family-free NEO orbit distribution model that also incorporated the observational selection effects typical of the asteroid surveys contributing the current NEO inventory.  The selection effects are important because \eg\ they favor the discovery of objects on Earth-like orbits and therefore increase the orbital element phase-space density of known objects on these types of orbits.

\subsubsection{Synthetic family-free NEO model}
\label{sec:syntheticNEOModel}

To generate our synthetic family-free NEO model we started with the set of NEOs from the Synthetic Solar System Model \citep[S3M;][]{gra2011}. The S3M includes over 11 million objects ranging from those that orbit the Sun entirely interior to the Earth's orbit to the most distant reaches of the solar system.  The 268,896 NEOs with absolute magnitude $H<25.0$ in the S3M were generated in accordance with the $(a, e, i)$ orbital element residence-time distribution of the \citet{bot2002a} NEO model.  The model does not include any NEO families. 

To model the observational selection effects on the S3M NEO population
we performed a long-term survey simulation using the Pan-STARRS Moving
Object Processing System \citep{den2007}.  MOPS was developed to
process source detection data from Pan-STARRS \citep{kai2010} but also
incorporates real-time processing of synthetic detections to monitor
the system's performance.  Thus, it can be also used as a pure survey
simulator.  

In an ecliptic longitude ($\lambda_0$) and latitude ($\beta$) system
centered on the opposition point $(\lambda_0,\beta)=(0\deg,0\deg)$ the
MOPS simulated survey\footnote{The simulated survey discussed herein
  is only loosely related to the actual PS1 survey.  The details of
  the survey simulation are not important --- all that matters is that
  the simulated survey reproduces the observational selection effects
  of the ensemble of surveys that produced the known NEO population.
  The litmus test is whether the resulting simulated orbit element
  distributions match the known orbit distributions.}  region is
broken into two regions covering about 5,500 deg$^2$: 1) the
opposition region with $|\lambda_0|<30\deg$ and $|\beta|<40\deg$ and 2)
two `sweet spots' with $|\beta|<10\deg$ and $60\deg<|\lambda_0|<90\deg$.

MOPS uses a full $N$-body ephemeris determination to calculate the
exact ($RA$, $Dec$) of every NEO in each synthetic field and then
degrades the astrometry to the realized PS1 astrometric error level of
0.1$\arcsec$ (\citet{mil2012} have shown that PS1 achieves $\sim
0.13\arcsec$ absolute astrometric error). The photometry for each
object is degraded in a S/N-dependent manner such that as
S/N$\rightarrow$5 the magnitude error approaches $\sim$0.1 mag.  MOPS
then makes a cut at S/N=5 to simulate the statistical loss of
detections near the PS1 system's limiting magnitude$^1$ of
$R\sim$22.7. Each field is observed twice each night within $\sim$15
minutes to allow the formation of tracklets, pairs of detections at
nearly the same spatial location, that might represent the same solar
system object. Fields are re-observed 3 times per lunation (simulated
weather permitting) and tracklets are linked across nights to form
tracks that are then tested for consistency using an initial orbit
determination (IOD).

%\begin{center}
%{\bf Fig.~\ref{fig:real-synthetic-fuzzy-neo-aei}}
%\end{center}

\begin{figure}
  \centering
  \includegraphics[width=0.8\textwidth]{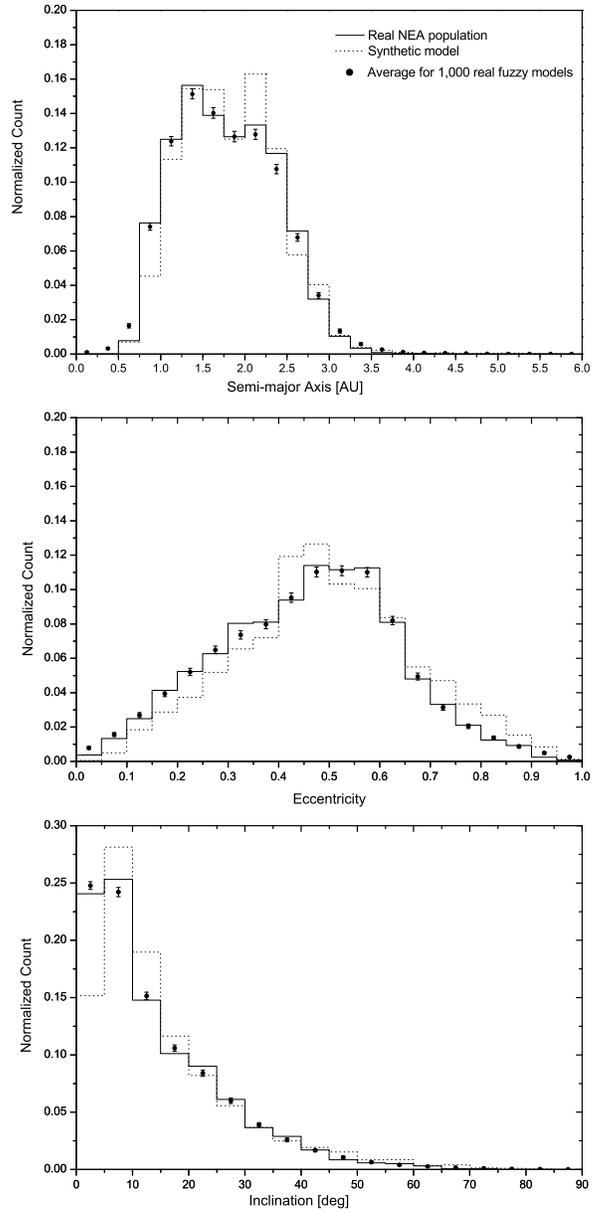}
  \caption[]{: Semi-major axis, eccentricity and inclination distributions of the real NEO population (solid line), the synthetic NEO model (dotted line) and our fuzzy-real NEO model (data points).} \label{fig:real-synthetic-fuzzy-neo-aei}
\end{figure}

Detections in tracks with small astrometric residuals in the IOD are
subsequently differentially corrected to obtain a final orbit.  Our
four-year MOPS survey simulation on the S3M NEO population yielded
8,020 derived objects (synthetic discoveries).
Fig. \ref{fig:real-synthetic-fuzzy-neo-aei} shows that the derived
synthetic NEOs are a rough match to the known NEOs.  Thus, our NEO
survey simulation yields a set of synthetic NEOs that are a proxy for
the known population including their observational selection effects
\citep{jed2002}. Pan-STARRS is currently operating a single prototype telescope on Haleakala, Hawaii, known as PS1.

\subsubsection{Cluster identification in the synthetic NEO model}
\label{sec:syntheticNEOModelClusterIdentification}

With a large amount of computing time we could identify the set of clusters detectable in the real NEO population for all possible combinations of the four thresholds --- \Dc, \Dp, $SCR_{min}$ and $PR_{min}$.  But this is impractical because we will also need to run orders of magnitude more tests to establish the detected clusters' statistical significance.  We were guided in our selection of the thresholds by previous work, our own experience, and testing the algorithm on the synthetic NEO model.

%\begin{center}
%{\bf Fig.~\ref{fig:SCR.vs.PF-synthetic}}
%\end{center}

\begin{figure}
  \centering
  \includegraphics[width=1.0\textwidth]{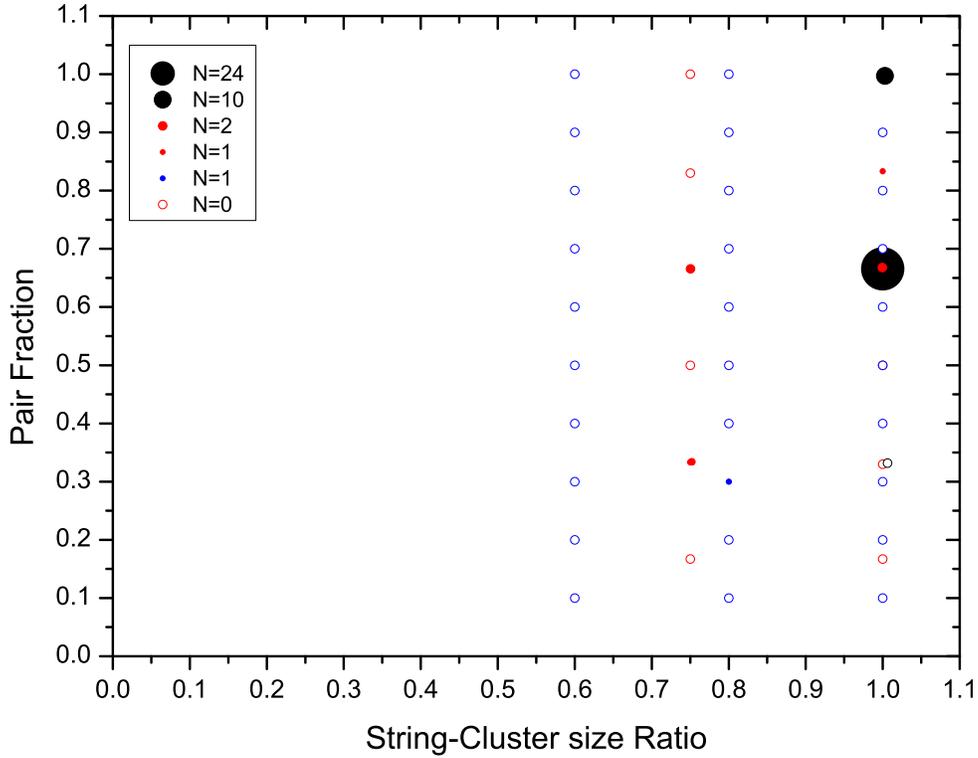}
  \caption[]{: Pair fraction ($PF$) vs. String to Cluster size Ratio ($SCR$) for false clusters detected in the family-free synthetic NEO model when $D_{cluster}=0.060$ and $D_{pair}=0.058$.  The area of the filled circles is proportional to the number of false clusters at each $PF$-$SCR$ combination.  Black circles represent clusters with 3 members, red is for clusters with 4 members, and blue represents clusters with $\ge$5 members.  Empty black, red and blue circles represent the quantized possible values for 3, 4 and $\ge$5 member clusters respectively. It is impossible to have the combinations of $PF$ and $SCR$ on the left side of the figure.}\label{fig:SCR.vs.PF-synthetic}
\end{figure}

Fig.~\ref{fig:SCR.vs.PF-synthetic} shows the results of the cluster
identification method applied to the family-free synthetic NEO model
using $D_{cluster}=0.060$ and $D_{pair}=0.058$ without any cuts on
$SCR$ and $PF$.  Even though our \DSH\ thresholds are already
considerably tighter than those proposed by \citet{fu2005} we still
identify many false clusters containing 3 members.  We even identify
one false 5-member cluster with $SCR=0.8$ and $PF=0.3$!  However, the
number of false clusters drops quickly with the number of cluster
members (42 triplets, seven 4-member clusters and only one 5-member
cluster).  As the synthetic model is a good representation of the real
NEO population we expect roughly the same number of false clusters
amongst the real NEOs when we use the same \Dc\ - \Dp\ cuts.
Thus, even though we will explore the full range of tighter thresholds
on the \Dc\ and \Dp\ values as described below, based on
Fig.~\ref{fig:SCR.vs.PF-synthetic} we will use $SCR_{min}=0.75$ and
$PF_{min}=0.5$ to identify clusters containing $\ge 4$ members and not
inspect the clusters containing NEO pairs and triplets.  Establishing
the smaller clusters' statistical significance will be difficult or
impossible using the techniques developed here.

In the following section \S\ref{sec:RealNEOClusters} we will use the
thresholds derived here to identify real NEO families for detailed
analysis.  However, when we measure the statistical significance of
those NEO families in \S\ref{sec:results} we will abandon the
synthetic model in favor of a more realistic one --- but the more
realistic model must be created using the thresholds derived here.

\subsection{NEO clusters in the real population}
\label{sec:RealNEOClusters}

We identified three clusters of four or more members in the real NEO
population from the mpcorb.dat database\footnote{\tt
  http://www.minorplanetcenter.net/iau/MPCORB.html}.  (We will ignore the 13 triplets and 243 pairs identified with
the same cuts.)  The members' orbital elements and other physical
parameters are provided in Tables~\ref{tab:NEOClusterOrbitElements}
and \ref{tab:NEOClusterPhysicalParameters}.  The three clusters are
labelled C1, C2 and C3, and have 4, 6 and 5 members respectively.  The
absolute magnitudes of the cluster members spans $18.5<H<29.5$
corresponding to diameters ranging from several meters up to several
hundreds of meters depending on the choice of albedo.

%\begin{center}
%{\bf Tab.~\ref{tab:NEOClusterOrbitElements}}
%\end{center}

The total number of $\ge 4$ member clusters identified in the real data
should be compared to the 0 $\ge 4$ member clusters identified in the
family-free synthetic data with $SCR_{min}=0.75$ and $PF_{min}=0.5$
(see Fig.~\ref{fig:SCR.vs.PF-synthetic}).  The disparity in the number
of $\ge 4$-member clusters could be due to the presence of real NEO
families but we will argue below that it is due to the lack of
fidelity in the synthetic NEO model --- we find that the synthetic NEO
model must be exquisitely matched to the real NEO population in order
to assess the statistical significance of the detected clusters.

The C1 cluster is the only one composed of objects with $H < 20$ and,
more importantly, the only one containing objects with orbital arcs
longer than 100 days.  The C1 cluster members' orbital element
uncertainties are typically 1-2 orders of magnitude lower than the other two clusters.  
All the C1 members belong to the Amor\footnote{The Amors
  have perihelion distance $q$ in the range 1.0167~AU$<q\le$1.3~AU,
  Apollos have $a>1.0$~AU and $q\le1.0167$~AU and the Atens have
  $a<1.0$~AU and aphelion $Q>0.983$AU.} NEO sub-population so
that they do not cross the Earth's orbit.  Indeed, the members of this
cluster have a perihelion distance of $\sim1.25$~AU implying that if
the cluster is a genetic NEO family it has probably never approached
close to the Earth or Venus.  Considering the similarity of all 5
orbital elements the cluster must have formed relatively recently and
it is unlikely that the cluster or its members could have approached
the Earth and then evolved onto orbits that do not cross the Earth's
on a short time scale.

In contrast to the C1 cluster, the C2 and C3 clusters are composed of
small objects with $H>21.1$ and $H>27.7$ respectively and include
objects with orbital arcs sometimes spanning just several days.  The
short arc lengths yield large uncertainties on the orbital elements
which in turn induce a large uncertainty in the C2 and C3 \DSH.  The
clusters thus illustrate how false associations can arise because of
the orbit element uncertainties.  The nominal orbits of the two
clusters place them in the Apollo NEO sub-population with the C2
cluster lying close to the Amor-Apollo transition and the C3 cluster
close to the Apollo-Aten transition.  Their location near the
transition regions is not a coincidence --- these small objects were
identified by NEO surveys only because their orbits bring them very
close to Earth.  The perihelion distance is $\sim1.00$~AU for the
members of the C2 cluster while the members of the C3 cluster are on
very Earth-like orbits with $a\sim1.00$ and $e<0.1$.  Establishing the
C2 and C3 clusters' statistical significance would be difficult
because observational selection effects are not well-characterized for
objects in their size range and this induces a large uncertainty in
the orbit and size-distribution models for small NEOs.

%\begin{center}
%{\bf Tab.~\ref{tab:NEOClusterPhysicalParameters}}
%\end{center}

Given the problems with the C2 and C3 clusters we will concentrate on establishing the statistical significance of the C1 cluster.  One method of doing so is to run the cluster finding algorithm on many instances of the synthetic family-free NEO model described in \S\ref{sec:syntheticNEOModel}.  If we identify $\le 3$ false clusters of $\ge 4$ members in 1,000 realizations of the synthetic NEO model we could claim that C1 is statistically significant with $\ge 99.7$\% or $\ge 3 \sigma$ confidence.  We did not employ this technique because of our concerns with the use of the $\sim10$~year old \citet{bot2002a} NEO model that underestimates the number of Amor-type NEOs like the members of the C1 cluster.  There are currently 474 {\it known} Amors with $H<18$ (as of March 2011) compared to the \citet{bot2002a} prediction of $310\pm38$ --- a $>4\sigma$ difference between the real and synthetic NEO populations.  If the \citet{bot2002a} NEO model underestimates the number of Amor-type NEOs then it would imply that we will overestimate the C1 cluster's statistical significance.  Furthermore, the synthetic NEO model relies on a survey simulation that was not intended to perfectly model real surveys and yields $\sim$6\% more objects than the real NEO population with small but perhaps significant skewing in the synthetic orbital element distributions as shown in Fig.~\ref{fig:real-synthetic-fuzzy-neo-aei}.   We need a better synthetic NEO model as described in the next section.

\subsubsection{Fuzzy real NEO models}
\label{sec:fuzzymodelproduction}

\begin{figure}
  \centering
  \includegraphics[width=\textwidth]{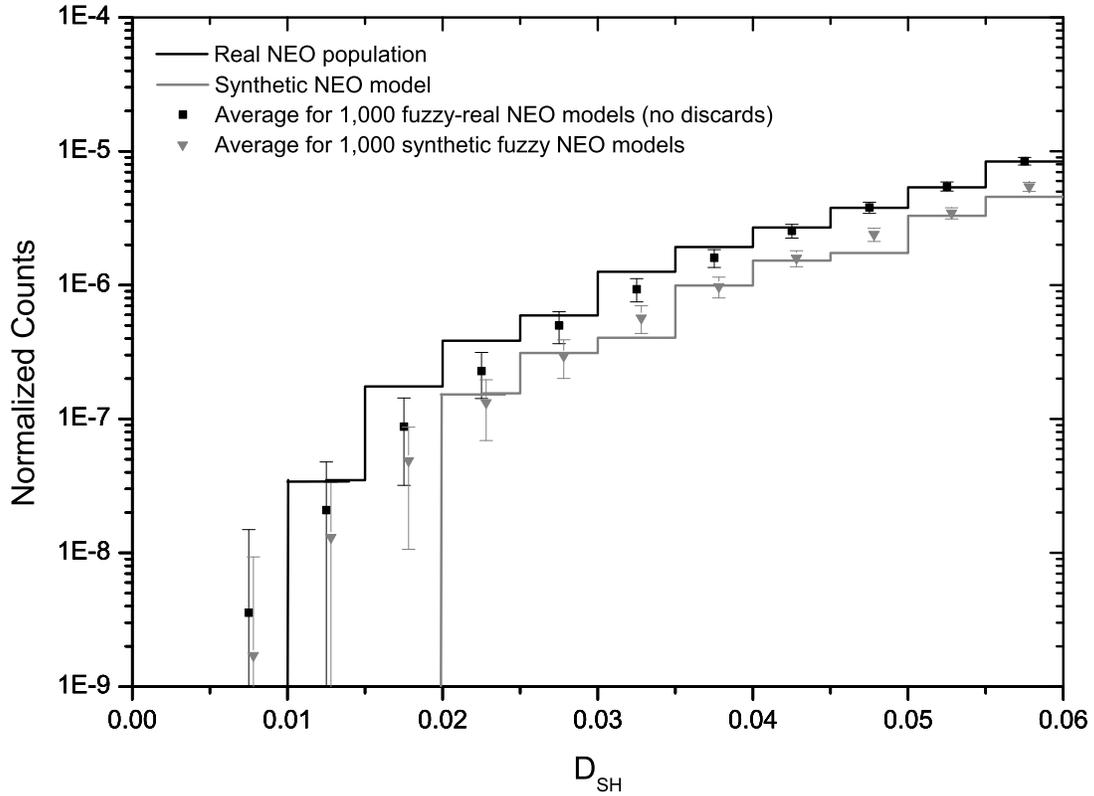}
  \caption[]{: $D_{SH}$ distribution of close NEO pairs normalized by the total number of possible pairs in each population.  The solid black histogram represents the real NEO population while the black data points represent the average$\pm$RMS of 1,000 `fuzzy' real NEO models (the `fuzzing' technique is described in \S\ref{sec:fuzzymodelproduction}).  The gray histogram represents the nominal synthetic NEO population while the grey data points represent the average$\pm$RMS of 1,000 `fuzzy' synthetic NEO models.}\label{fig:DSH-for-real-and-synthetic-close-pairs}
\end{figure}

The main problem with the \citet{bot2002a} synthetic NEO model is
illustrated in Fig.~\ref{fig:DSH-for-real-and-synthetic-close-pairs}
--- there is a huge discrepancy between the normalized
\DSH\ distributions for the closest pairs within the real and synthetic NEO populations.  But establishing the statistical
significance of our NEO clusters requires a large number of
independent high-fidelity family-free NEO models that incorporate observational
selection effects.  Thus, we developed NEO models using a technique
that i) maintains both the micro and macro distribution of 5 orbital
elements (ignoring the mean anomaly) and ii) eliminates any possible
real clusters.

%\begin{center}
%{\bf Fig.~\ref{fig:DSH-for-real-and-synthetic-close-pairs}}
%\end{center}

Our solution was to `fuzz' the orbital elements of each real NEO
around its position in 5-dimensional orbital element space in a manner
that maintained the local orbital element phase-space density and thereby
preserves both the intrinsic NEO orbital element distribution and the
observational selection effects.  For each NEO ($k$) in the real
population we identified its closest neighbor as the object with the
smallest $D_{SH} \equiv D_{fuzz}$.  We then generated a new `fuzzy'
synthetic orbit ($n$) that has $D_{SH} \le D_{fuzz}$ with respect to
the original orbit as described below.

If all the difference between the original and new orbit is due to a single orbital element (\eg\ $\Delta a = a_n - a_k$) then we obtain $\Delta a$, $\Delta e$, $\Delta i$, $\Delta \omega$, and $\Delta \Omega$ from:
\begin{eqnarray}
 D_{fuzz} &=& \frac{(\Delta a)}{AU}(1-e) \\
          &=& \Delta e \; \sqrt{1 + (a/AU)^2} \\
          &=& 2 \; \sin(\Delta i/2) \\
          &=& 2 e \; \sin(\Delta \omega / 2) \\
          &=& \sqrt{{{d_3}^\prime}^2+{{d_4}^\prime}^2} \\
\end{eqnarray}
where
\begin{eqnarray}
 {d_3}^\prime &=& 2 \sin{(I_{\Delta \Omega}/2)},  \\
 {d_4}^\prime &=& 2e \; \sin({\Pi_{\Delta \Omega}/2}),
\end{eqnarray}
and
\begin{eqnarray}
  I_{\Delta \Omega} &=&
\arccos{\left[\cos^2{i}+\sin^2{i}\cos{\Delta \Omega}\right]} \\
 \Pi_{\Delta \Omega} &=& 2\arcsin{\left[\cos{i} \sin\frac{\Delta
\Omega}{2}\sec{\frac{I_{\Delta \Omega}}{2}}\right]}.
\end{eqnarray}
We then
generated a `fuzzed' orbit with $(a',e',i',\omega',\Omega')$ where the
elements $x'$ were generated randomly within the range
[$x-\Delta x,x+\Delta x$]. Finally, we calculated the $D_\Delta$
between the original and the fuzzed orbit and repeated the generation of
the fuzzed orbit for the object until $D_\Delta <D _{fuzz}$.   We also repeated the synthetic object generation if the new orbit was not a NEO (had perihelion $a'(1-e')>1.3$~AU), was hyperbolic (\ie\ $e'\ge1$), or unphysical (\eg\ $a'\le0$, $e'\le0$).  We generated 1,000 instances of these `fuzzy-real NEO models' to be used for establishing the statistical significance of our NEO clusters.

We tested the generation of the fuzzy-real NEO models by generating a series of models fuzzed by $D_{fuzz}' = f_{fuzz} \; D_{fuzz}$ with $f_{fuzz}=0.0$, 0.2, 0.5, 1.0, 2.0, and 4.0 and verified that the models behave as expected and as $f_{fuzz}\rightarrow 0$ the generated model reproduces the input model exactly.  

The remaining problem with the fuzzy-real NEO models is that if the
real NEO population contains real families then so will the fuzzy-real
NEO models.  We needed to remove any real NEO families from the model
first --- but this is difficult to accomplish when there are no known
real NEO families.  Instead, we used our own cluster results
agnostically with the assumption that it does not matter whether the
clusters we identified are real or not, all that matters is how often
the fuzzing process generates false clusters.  Thus, {\tt i}) using
the cluster detection thresholds determined with the synthetic NEO
population described in \S\ref{sec:syntheticNEOModel} we identified
all clusters containing $\ge 3$ members and {\tt ii}) treated the
largest member of each cluster as any other NEO as described above but
{\tt iii}) fuzzed the orbits of the 18 smaller members of the clusters
with $f_{fuzz}=10$.  We needed to keep the 18 small objects in the
model because they represent about 0.3\% of the total known NEO
population --- about equal to the 3$\sigma$ contribution to the NEO
model that we were attempting to measure.  On the other hand we needed
to keep them in roughly the correct location in the NEO orbit
distribution.

%\begin{center}
%{\bf Fig.~\ref{fig:D.nearestneighbors.real.vs.fuzzed}}
%\end{center}

\begin{figure}
  \centering
  \includegraphics[width=\textwidth]{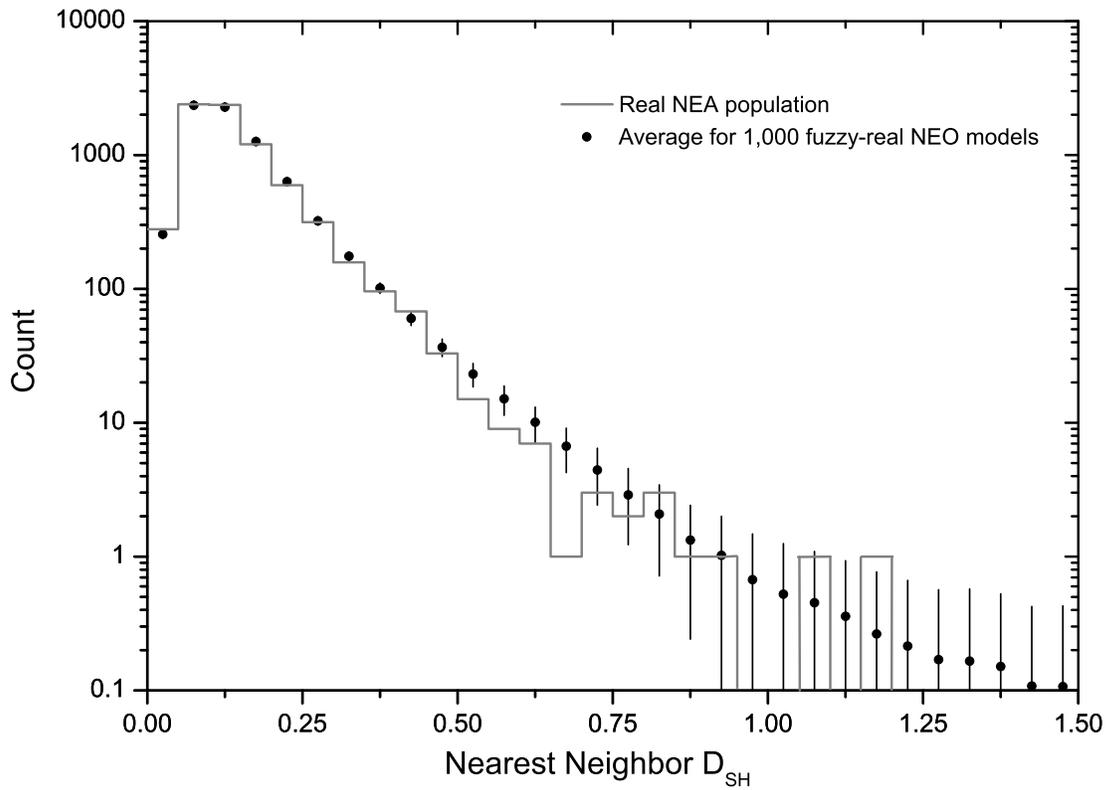}
\caption[]{: \DSH\ distribution for the closest neighboring objects in the real NEO population and the average$\pm$RMS for 1,000 fuzzy-real NEO models.}\label{fig:D.nearestneighbors.real.vs.fuzzed}
\end{figure}

Fig.~\ref{fig:real-synthetic-fuzzy-neo-aei} shows that the fuzzy-real
NEO models' semi-major axis, eccentricity and inclination
distributions match the known NEO population far better than the
synthetic NEO model.  Even more importantly,
Figs.~\ref{fig:DSH-for-real-and-synthetic-close-pairs} and
\ref{fig:D.nearestneighbors.real.vs.fuzzed} show that the fuzzy-real
NEO models preserve both the micro and macro \DSH\ distributions
(respectively) of the real NEO population that are critical to using
the models to establish the statistical significance of the NEO
clusters.

Note that the \DSH\ distribution of the real NEOs is systematically
slightly higher than the fuzzy-real NEO model at small \DSH\ in
Fig.~\ref{fig:DSH-for-real-and-synthetic-close-pairs}.  This would be
the expected signature if there were real families in the real NEO
population.  The data point in the lowest \DSH\ bins in
Fig.~\ref{fig:DSH-for-real-and-synthetic-close-pairs} represent two
separate pairs of NEOs.  If these two pairs were real it would imply
that they were statistically significant but it turned out that they
were subsequently identified by the Minor Planet Center as
corresponding to the same physical object.  Thus, it is reassuring 1)
that our fuzzy-real NEO model \DSH\ distribution agrees well with the
real NEOs and 2) that our technique successfully identified identical
NEOs.

\section{Results and discussion}
\label{sec:results}

\begin{figure}
 \centering
 \includegraphics[width=\textwidth]{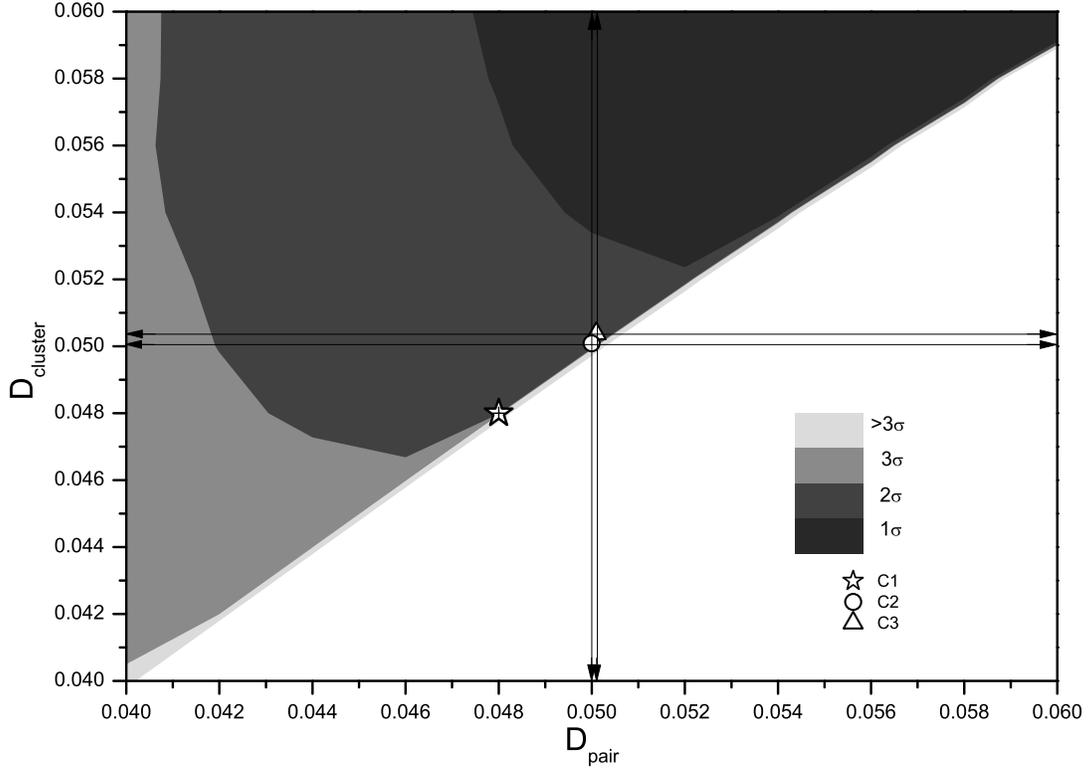}
  \caption[]{: Statistical significance of the C1, C2 and C3 clusters as a function of the \Dc\ and \Dp\ thresholds with $SCR_{min}=0.75$ and $PF_{min}=0.5$. The shaded grey regions mark areas of 3-, 2-, and 1-$\sigma$ significance for clusters with $\ge 4$ members using the fuzzy-real NEO models (see text for details). The star, circle and triangle represent the most statistically significant location of the C1, C2 and C3 clusters respectively.  The small black lines on the C1 data point represent the derived errors in \Dc\ and \Dp\ due to the orbital element errors.  The long vertical and horizontal lines through the C2 and C3 cluster's data points represent the same errors for those clusters. The white region under the diagonal line is unphysical with $D_{cluster} < D_{pair}$.}\label{fig:1000.RealNEO.Fuzz.Models.Contourplot}
\end{figure}

The shaded regions in
Fig.~\ref{fig:1000.RealNEO.Fuzz.Models.Contourplot} show the results
of running our cluster identification algorithm on the 1,000
family-free fuzzy-real NEO models described above.  We searched for
clusters containing $\ge 4$ members over the range of \Dc--\Dp\ space
with $0.040 \le D_{cluster} \le 0.060$ where $D_{pair} \le
D_{cluster}$ using fixed $SCR_{min}=0.75$ and $PF_{min}=0.5$ as
described in \S\ref{sec:syntheticNEOModelClusterIdentification}.  It
is surprisingly easy to create NEO clusters with $\ge 4$ members. The
shaded areas in the figure correspond to regions where we detected a total of a minimum of 3, 46 or 317 clusters in the 1,000 models.  The regions correspond to 3-, 2- and
1$\sigma$ confidence levels on the detection of clusters in the real
NEO population.

%\begin{center}
%{\bf Fig.~\ref{fig:1000.RealNEO.Fuzz.Models.Contourplot}}
%\end{center}

Fig.~\ref{fig:1000.RealNEO.Fuzz.Models.Contourplot} also shows the
location of our detected C1, C2 and C3 clusters in the real NEO
population.  The clusters can be identified over a limited range of
\Dc--\Dp\ with the most statistically significant point typically
having the smallest \Dc\ and \Dp.  It may be surprising that the
clusters are {\it not} identified in a broad fan extending upwards and
to the right in the figure.  The truncation in the region in which
they are identifiable is due to the $SCR_{min}=0.75$ and
$PF_{min}=0.5$ cuts --- using looser values of \Dc\ and \Dp\ allows
the cluster to `absorb' other nearby objects but these objects then
typically drive the cluster's $SCR$ and $PF$ values below the
$SCR_{min}$ and $PF_{min}$ thresholds.

Fig.~\ref{fig:1000.RealNEO.Fuzz.Models.Contourplot} shows that the
C1 cluster is statistically significant at $\sim$2.0$\sigma$.  The
C2 and C3 clusters are likely even less significant based on their
large \Dc\ and \Dp\ error bars extending well into the insignificant
region of the figure.  It is important to remember that our calculated
statistical significance of the NEO clusters hinges on the reliability
of our fuzzy-real NEO models described in
\S\ref{sec:fuzzymodelproduction}.  

Given that all 4 C1 members are relatively large NEOs with $H<20$ we
repeated the entire search for clusters using only those NEOs with
$H<20$ and re-calculated their statistical significance.  As expected,
the statistical significance of the C1 cluster increases to about 3$\sigma$ because of the reduction in the total number of NEOs to
$\sim 3100.$  

While we have taken every
precaution in developing our NEO models we understand that they have
their limitations.  For instance, the statistical significance of the
NEO clusters increases/decreases if we `fuzz' the model more/less
($f_{fuzz}>$/$<$1).  We assume that the `natural' value is
$f_{fuzz}=1$ and use this value when quoting the statistical
significance of our NEO clusters.  In the remainder of this work
we consider only the C1 cluster as a candidate NEO family.

\subsection{Dismembering the C1 cluster}
\label{ss.C1cluster}

In this sub-section we perform several tests of the C1 cluster to
determine if it is consistent with being a genetically related NEO
family.

\subsubsection{Provenance and taxonomy}
\label{sss:ProvenanceAndTaxonomy}

Despite our minor caveats with the \citet{bot2002a} NEO model
described above it can still be used to estimate the probability that
a NEO derives from one of their five NEO source regions.  The C1
cluster falls in the NEO model bin with central $a=2.150$~AU,
$e=0.425$ and $i=7.5\deg$ for which objects have $\sim$35\%
probability of deriving from the $\nu_6$ region and a $\sim$63\%
chance of having evolved from the so called `Mars Crossing' region.
In other words, a nearly 100\% probability of deriving from a source
in the inner region of the main belt.

The inner region of the belt is dominated by S-class asteroids
\citep[\eg][]{zel1979} so it is a little surprising that SDSS
spectrophotometry \citep[from the 4$^{th}$ Moving Object Catalog,
  MOC4;][]{izv2002b} for the only available C1 cluster member,
2000~HW$_{23}$, is not S-like\footnote{We used the solar colors from \citet{bilir2005,rodgers2005}}.  We performed a simple linear
interpolation and extrapolation of the 2000~HW$_{23}$ SDSS MOC4 5-band
spectrophotometry and error bars to the SMASSII bands' central
wavelengths \citep[Phase II of the Small Main-Belt Asteroid
  Spectroscopic Survey;][]{bus2002}.  (The only extrapolation was from
the real data point at 892~nm to the data point at 920~nm.)

\begin{figure}
  \centering
  \includegraphics[width=\textwidth]{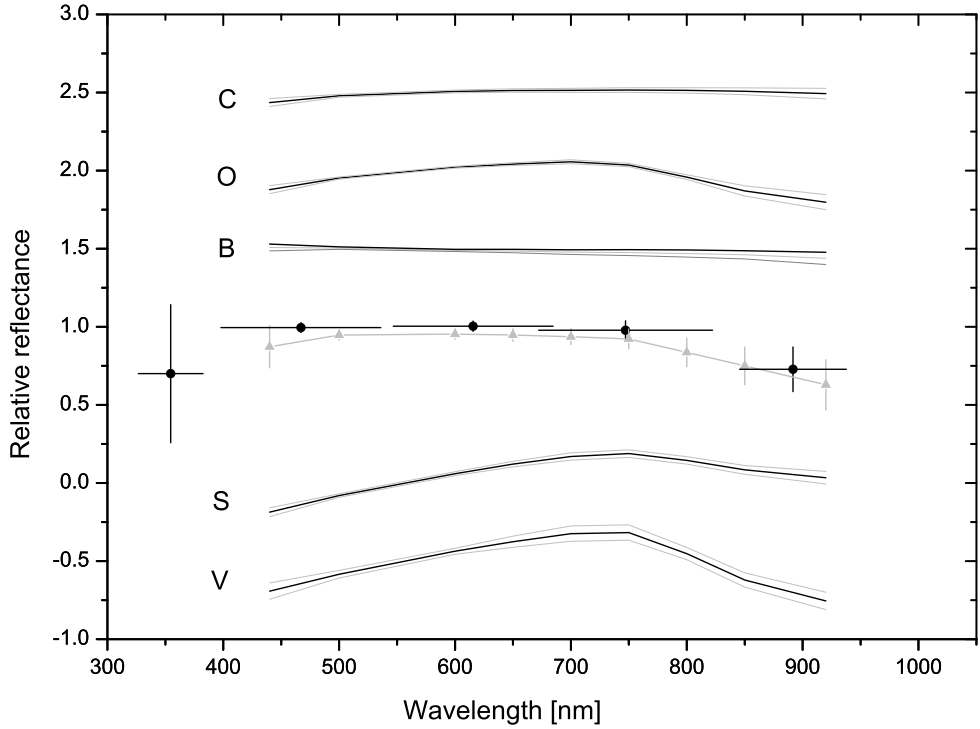}
 \caption[]{: Spectrophotometry of the C1 cluster's largest member,
   2000~HW$_{23}$, from the SDSS MOC4 by \citet{izv2002b} (black
   points) and the corresponding interpolated and extrapolated data
   points and error bars at the SMASSII filters' central
   wavelengths \citet{bus2002} (grey triangles and connecting
   lines).  The synthetic SMASSII data points and lines are offset by
   -0.02 for clarity. The thin black curves and their adjacent grey
   curves represent the mean$\pm$RMS respectively of 5 asteroid
   taxonomic types from the SMASSII survey.  All distributions have
   been corrected to solar colors and normalized to 1.0 at 550~nm.
   The relative reflectances of the C, O, B, S and V types are offset
   by +1.5, +1.0, +0.5, -1.0 and -1.5
   respectively.}\label{fig:2000HW23-reflectance-vs-wavelength}
\end{figure}

Even if at first glance the asteroid appears to be most similar to the
O-type due to the dip in reflectivity at long wavelengths (see
Fig.~\ref{fig:2000HW23-reflectance-vs-wavelength}) a normalized
$\chi^2$ fit of the 2000~HW$_{23}$ spectrophotometry to 26 major
asteroid classes from $A$ through $X_k$ \citep[from][]{bus2002}
suggests that it is most consistent with the unusual B-type, a subset
of the larger C-complex.  The O-type provides a close second best fit.
In general, keeping in mind that the fit used data extrapolated from
the SDSS MOC4 catalogue, 2000~HW$_{23}$ yields good fits with the
different types in the C-complex --- not the S-class that dominates
the inner main belt.

%\begin{center}
%{\bf Fig.~\ref{fig:2000HW23-reflectance-vs-wavelength}}
%\end{center}

The case for the C1 cluster being a legitimate family will be
strengthened if the spectra of other C1 cluster members can be shown
to be similar to 2000~HW$_{23}$ and belonging to the C-complex.

\subsubsection{Size-frequency distribution}

Under the assumption that C1 is a family we can estimate the slope of
the family's size-frequency distribution (SFD) assuming that the SFD
is proportional to $10^{\alpha H}$.  An unreasonable value of the
slope would indicate that the cluster might not be a genetic family.
Converting the observed C1 cluster member's $H$ values into the true
SFD requires the detection efficiency as a function of absolute magnitude, $\epsilon(H)$,
for objects with C1-like orbits.  We estimated $\epsilon(H)$ using the
synthetic survey simulation described in \S\ref{sec:syntheticNEOModel}
by dividing the number of derived objects with C1-like orbits by the
number of objects in the input model.  The survey simulation does a
good enough job of reproducing the real NEO population for the purpose
of estimating this efficiency and will not dominate the induced error
on the measured SFD.  We found a reasonable fit to the efficiency from
the synthetic data using
\begin{equation}
\epsilon(H) = { {\epsilon_0} \over { 1 + \exp[(H-L)/w] } }
\label{e:C1DetectionEfficiency}
\end{equation}
with nominal values of $\epsilon_0=1.0$, $L$=20.0 and $w=0.76$. A
maximum likelihood fit to the 4 C1 members' $H$ distribution yields
$\alpha=0.23^{+0.10}_{-0.08}$ where the error bars are
statistical-only.  The statistical errors on the fit are much larger
than the systematic errors introduced by the uncertainty on the
efficiency function parameters.

\begin{figure}
  \centering
  \includegraphics[width=\textwidth]{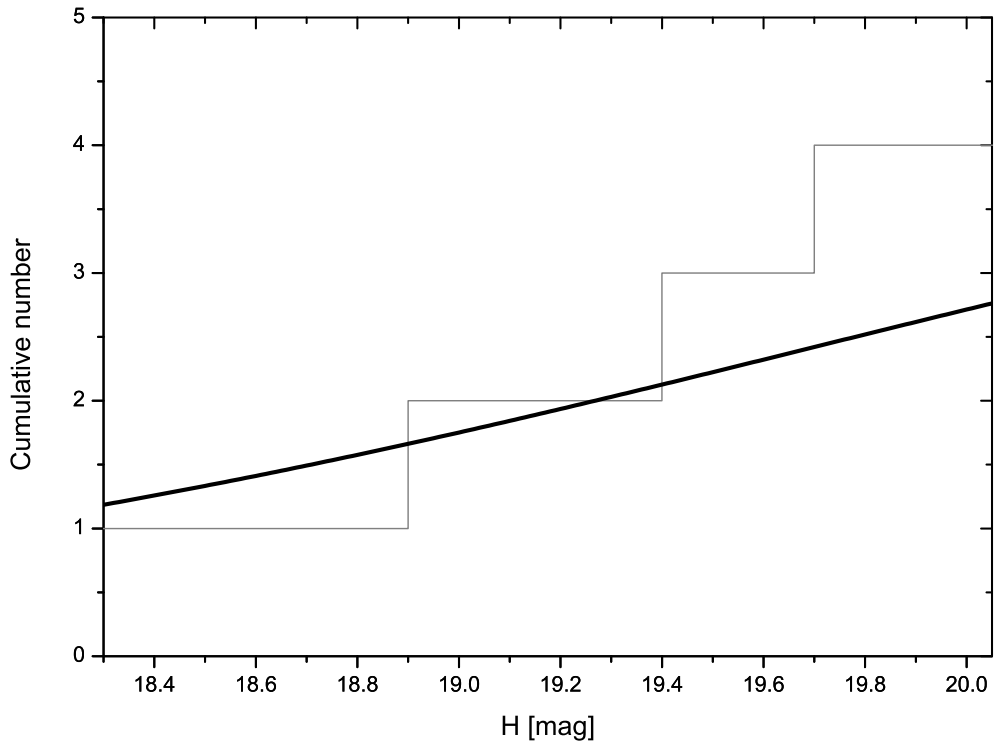}
 \caption[]{: Cumulative absolute magnitude distribution of the four C1
   cluster members (grey).  The thick black curve represents the
   product of the efficiency function (see text) with the result of
   the maximum likelihood fit to the SFD with
   $\alpha=0.23^{+0.10}_{-0.08}$.}\label{fig:C1-SFD}
\end{figure}

The slope of the C1 cluster's SFD (Fig.~\ref{fig:C1-SFD}) is shallower but not inconsistent
with the overall NEO population's SFD of 0.35$\pm$0.02 measured by
\citet{bot2002a} for $H$<22 and it is surprisingly close to the value
of 0.26$\pm$0.03 for H$\gtrsim$18 from \citet{mainzer2011}. \citet{parker2008} measured the SFD for
many main belt families and found that they can have complicated SFDs
that can not be fit by a single power-law and that they have a wide
range of slopes.  The slope varies from 0.35 to 0.97 for single-slope
families and from 0.10 to 0.62 at the faint end of the $H$
distribution for families with a broken power law SFD (\ie\ $H$ values
approaching those of the C1 cluster members).  Furthermore,
\citet{rich1998} suggests that tidally disrupted asteroid families
have shallow SFDs. Thus, we can not use the SFD slope to exclude the
hypothesis that the cluster is a family since the measured value is
consistent with slopes of established families.

%\begin{center}
%{\bf Fig.~\ref{fig:C1-SFD}}
%\end{center}

\subsubsection{Backward orbit integrations \& tidal disruption at Mars}

\begin{figure}
  \centering
  \includegraphics[scale=0.65]{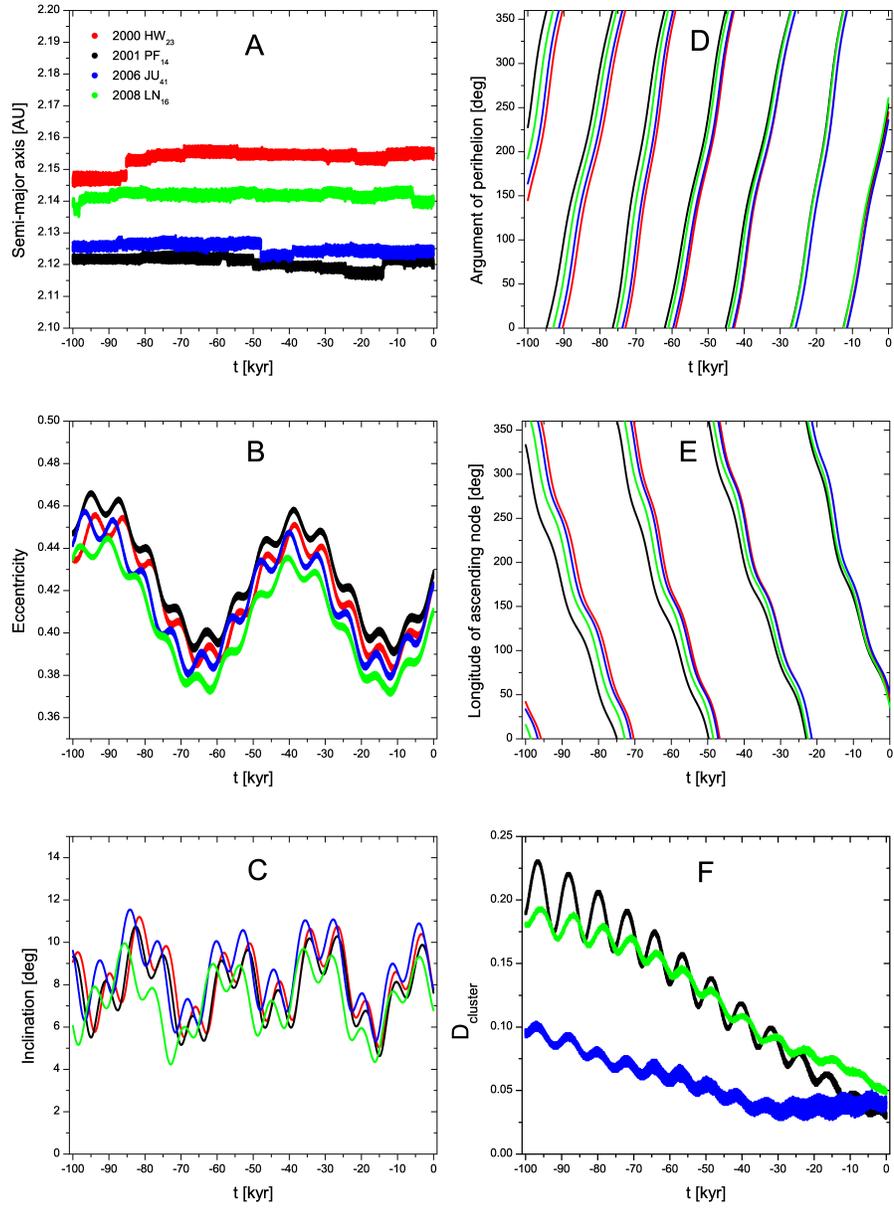}
 \caption[]{: Panels A-E: Backward evolution of the C1 cluster members'  orbital elements from the present ($t_0$) to 100~ky in the past.  Panel F: $D_{cluster}$ evolution for the C1 cluster from $t_0$ to 100~ky in the past.  $D_{cluster}$ was calculated with respect to 2000~HW$_{23}$ and therefore it does not appear on the panel. $D_{cluster}$ with respect to 2000~HW$_{23}$ does not differ significantly from the value calculated with respect to the average  orbit of the C1 members.}\label{fig:C1-orb-evolution}
\end{figure}

If the C1 cluster is the remnant of a recently formed family then we
expect that a backward integration of the 4 member's nominal orbits
would show a rapid convergence to a single orbit and the time frame
for the convergence would indicate the time of the family's formation
\citep[\eg\ as observed in the Karin family;][]{nes2006b}.  On the
contrary, Fig.~\ref{fig:C1-orb-evolution} shows that the nominal
orbits are similar but undergo a gradual dispersal moving 100~ky into
the past.  \ie\ they appear to have converged to their most similar
orbits at the present time.  The semi-major axis, eccentricity and
inclinations of the members remain similar but the longitude of
perihelion and ascending nodes gradually diverge.  The effect is
summarized in panel Fig.~\ref{fig:C1-orb-evolution}F showing the
evolution of the cluster's \DSH\ with time.  We note that two pairs of
objects remain tight in their mutual \DSH\ for the entire integration.

Given the apparent stability of the cluster members' orbits we
integrated all 4 nominal orbits backwards for 10 million years and
found that they are exceptionally similar back to 1.5 Myr in the past.
The mutual orbital stability is unusual for NEOs given that the
numerical integrations showed numerous close encounters of all objects
with Mars during this time.  The lack of convergence of the nominal C1
members' orbital elements suggests that they are not genetically
linked.  To firmly establish the statistical significance of the C1
cluster's genetic relationship would require the generation of
hundreds of clones of each C1 member and similarly generating clones
for a large number of C1-like clusters.

%\begin{center}
%{\bf Fig.~\ref{fig:C1-orb-evolution}}
%\end{center}

Rather than explore that computationally challenging route we explored
a few instances of the opposite question --- is it possible that a
family producing event in the recent past can reproduce the observed
C1 cluster's orbit and \DSH\ distribution?   We created 500 clones for each of the four
C1 members where each clone is consistent with the available
astrometry \citep[the clones were generated using the covariance
  sampling technique in the OpenOrb orbit-computation
  package;][]{grn2009}.  We integrated all the clones backwards in
time for 100~ky and discovered that all of the objects suffered
several close approaches with Mars and a large number of clones
approached to within Mars' Hill sphere.  Since the C1 cluster is
composed of Amor type NEOs we did not expect nor identify any close
approaches to the Earth and Venus. 

One 2000~HW$_{23}$ clone passed inside Mars' Roche limit $\sim$70,800
years ago and is therefore a candidate for tidal disruptions (assuming
that the parent body is a spherical fluid or rubble pile with a
density $\rho$=1.95~g-cm$^{-3}$ like (25143)~Itokawa \citep{Abe2006},
Mars's Roche limit is $\sim$3.08~Mars radii).  This result needs to be
tempered with a comparison to the likelihood that {\it any} NEO will
approach to within Mars' Roche limit in the same time frame.  We
found that 5,070 of 8,133 known NEO's nominal orbits approach Mars to
within 1 Hill radii at least once during the last 100,000 years.
Assuming that close approaches to within the Roche limit are as likely for clones of
all NEOs (1 in 2000 C1 clones) then we might expect that about 2.5 of
the known NEOs have actually approached to within Mars' Roche limit
and could be candidates for tidal disruption.  Perhaps the C1
cluster's hypothetical parent body is one of them and we will
represent it as $\overline{\tt{2000~HW}_{23}}$ --- the clone that made
the closest approach to Mars mentioned above.

To study the properties of a family of objects created by tidal
disruption by Mars we created twelve $\overline{\tt{2000~HW}_{23}}$
clones at the moment of its closest approach of 0.63 Mars Roche radii.
The twelve secondary clones have $\Delta\vec{v_x}$, or
$\Delta\vec{v_y}$, or $\Delta\vec{v_z}$ of $\pm$1~m-s$^{-1}$ and
$\pm$5~m-s$^{-1}$ with respect to $\overline{\tt{2000~HW}_{23}}$.
Simulations suggest that the tidal disruption process creates
fragments with relative speeds of $<1$~m-s$^{-1}$ (Kevin Walsh,
personal communication) so that our integrations will overestimate
the spread in the orbital elements of tidally disrupted families.  All
13 objects ($\overline{\tt{2000~HW}_{23}}$ and its 12 secondary
clones) were integrated forward from the time of closest encounter
to the present epoch with the same Bulirsch-Stoer integration routine and same time step as our other
integrations.

\begin{figure}
  \centering
  \includegraphics[scale=0.65]{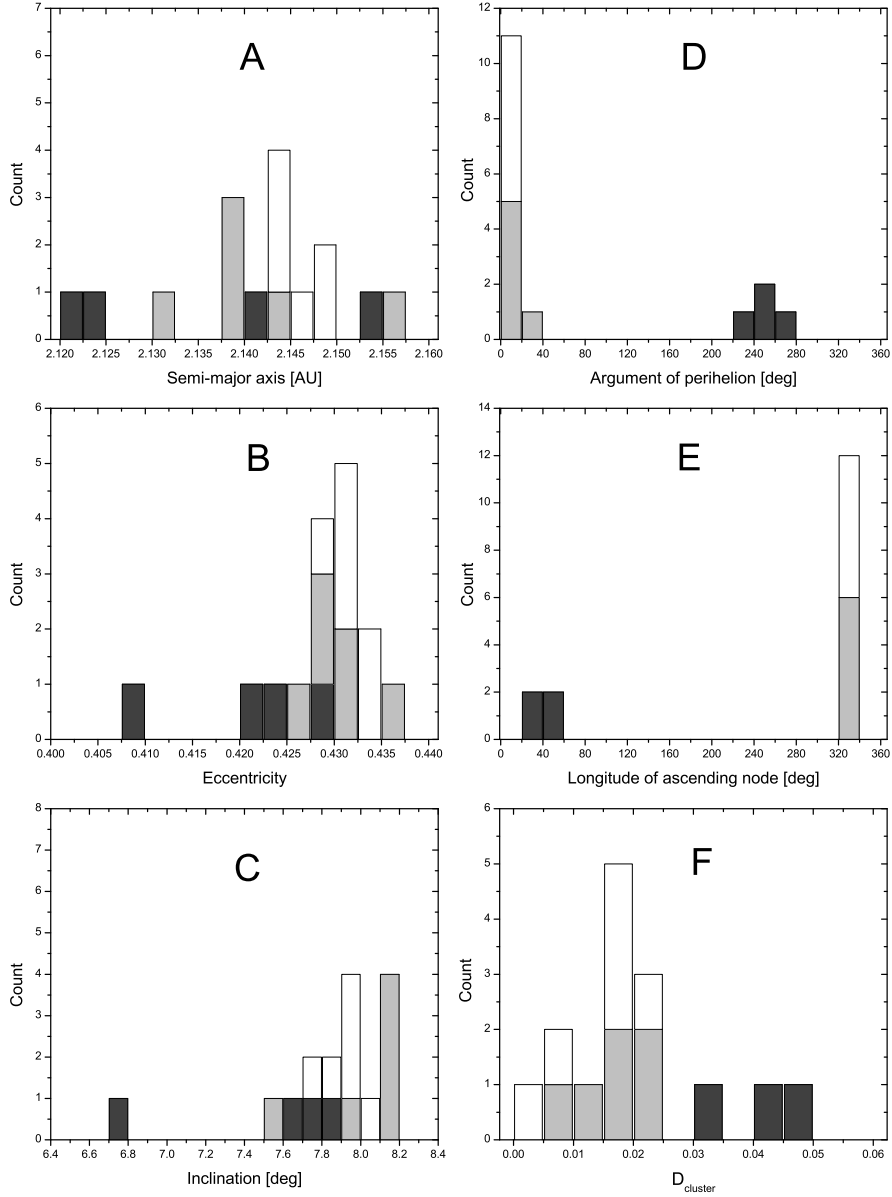}
\caption[]{: Panels A-F: Orbital elements and $D_{cluster}$ distribution for the C1 custer members (dark grey) and two sets of 6 secondary $\overline{\tt{2000~HW}_{23}}$ clones (see text for description). The two sets of clones have $\Delta\vec{v_x}$, or $\Delta\vec{v_y}$, or  $\Delta\vec{v_z}$ with respect to $\overline{\tt{2000~HW}_{23}}$ of $\pm$1~m-s$^{-1}$ (light grey) and $\pm$5~m-s$^{-1}$ (white) respectively at the time of closest Mars approach. There are only 3 values for the real C1 cluster because $D_{cluster}$ was calculated with respect to 2000~HW$_{23}$. The $D_{cluster}$ values for both sets of 6 secondary clones were calculated with respect to $\overline{\tt{2000~HW}_{23}}$ that was integrated simultaneously.}\label{fig:CE-clone-propagation}
\end{figure}

Panels A-E in Fig.~\ref{fig:CE-clone-propagation} show that the values
and ranges of $a$, $e$ and $i$ for the $\pm$1~m-s$^{-1}$ and
$\pm$5~m-s$^{-1}$ groups of secondary clones are comparable to the
values and ranges of the real C1 members.  The ranges of $\Omega$ and
$\omega$, $\ie$   the spread of values within the two sets of
$\overline{\tt{2000~HW}_{23}}$ clones and the 4 C1 cluster members,
are also similar.  The values of $\Omega$ and $\omega$ for the
$\overline{\tt{2000~HW}_{23}}$ clones and the 4 C1 cluster members do
not need to agree at the present time because the
$\overline{\tt{2000~HW}_{23}}$ secondary clones were created from a
single 2000~HW$_{23}$ clone whose orbit is slightly different from
2000~HW$_{23}$ due to the former's deep approach to Mars (their
semi-major axes differ by 0.05~AU).  The difference between $\Omega$
and $\omega$ for the $\overline{\tt{2000~HW}_{23}}$ clones and the 4
C1 cluster members is thus due to their different precession rates
that are sensitive to the initial values of semi-major axes \citep{murray1999}.
Note that the actual C1 cluster member's values show evidence of
slightly more spreading than the $\overline{\tt{2000~HW}_{23}}$
secondary clones.  Indeed, Fig.~\ref{fig:CE-clone-propagation}F shows
that the distribution of $D_{cluster}$ for the
$\overline{\tt{2000~HW}_{23}}$ secondary clones is also tighter than
the C1 members.  Both observations can be explained as being due to
non-gravitational forces acting on the real objects over the course of
the last $\sim$71~ky.

%\begin{center}
%{\bf Fig.~\ref{fig:CE-clone-propagation}}
%\end{center}

The distributions of orbital elements and $D_{cluster}$ in
Fig.~\ref{fig:CE-clone-propagation} suggests that the C1 cluster is
consistent with an origin in the tidal disruption of an asteroid
during a tidal disruption event with Mars 70,800 years ago.  If true,
the C1 cluster is a very young family.  In a followup paper to this
one we will show that the lifetime of tidally disrupted families
created by Mars are as long as $\sim$1~Myr so if they exist it may not
be surprising that we are beginning to detect them.  Furthermore, if
there is a tidally disrupted asteroid family in the NEO population it
seems more likely to be a C-class asteroid and, {\it ceteris paribus},
will disrupt more easily since they have a lower bulk density than the
S-class. \citep{rich1998}.  Thus, it is interesting that one C1
member with spectrophotometry, 2000~HW$_{23}$, may be in the C-class
(see \S\ref{sss:ProvenanceAndTaxonomy}).

\subsubsection{Proper element cluster search}

We also performed a cluster search with NEO proper elements
\citep{gro2001}.  Since the C1 cluster was detected as a cluster in
osculating element space it seems plausible that it should also be
detectable in proper element space as described in the introduction.
A complication arises because our cluster search algorithm uses a
5-element \DSH-criterion that incorporates the longitude of the
ascending nodes and argument of perihelion (the $d_3$ and $d_4$ terms
in eq.~\ref{eq:DSH}) while only the proper semi-major axis,
eccentricity and inclination ($a_{p}, e_{p}, i_{p}$) can be calculated
for asteroids.  Thus, the \DSH\ for clusters in proper element space,
$D_p$, must be smaller than those found using the 5-element \DSH\ with
the osculating elements ---  $D_{cluster}=0.002$ compared
to $D_{cluster}=0.040$ used for the osculating element cluster search.

The proper element cluster search revealed that the C1 cluster is not
outstanding among NEO clusters.  We identified a C1$_p$ cluster in the
proper element search that has 3 objects in common with the C1 cluster
--- 2000~HW$_{23}$, 2001~PF$_{14}$, and 2008~$LN_{16}$ --- and there
are 166 clusters with 5 or more members that are bound tighter than
C1$_p$ with $D_p$=0.018.  The C1$_p$ cluster contains two additional
objects, 2002~RA$_{182}$ and 2004~TE$_{18}$, instead of 2006~JU$_{41}$. The osculating element \DSH\  for
2004~TE$_{18}$ and 2002~RA$_{182}$ calculated with respect to
2000~HW$_{23}$ are 0.471 and 0.125, much too large to indicate an
orbital similarity. 

We found that the stability of the NEOs in the C1 and C1$_p$ clusters
is not unusual.  We computed the MOID with all the planets from
Mercury to Neptune during the secular evolution of 8,650 NEOs (NEODyS,
29/2/2012) according to the dynamical model used in the computation of
the proper elements \citep{gro2001}.  Roughly 5\% of all NEOs (429) do
not cross the trajectory of {\it any} planet and $\sim42$\% of them
(3,636) cross only Mars' trajectory during their secular evolution.

%\begin{center}
%{\bf Tab.~\ref{tab:minmoid}}
%\end{center}

The six asteroids in the C1 and C1$_p$ clusters are of the latter
type.  Table~\ref{tab:minmoid} shows that all the cluster's members
remain far from both the Earth and Jupiter, the major NEO perturbers,
during their entire secular cycles.  Of the total 3,636 Mars crossing
objects we find that
\begin{itemize}
\item $\sim$50\% of them have $\min$(MOID$_{Earth}$)   $\ge 0.1$ AU, and
                         $\min$(MOID$_{Jupiter}$) $\ge 1$ AU,
\item $\sim$16\% of them have $\min$(MOID$_{Earth})$   $\ge 0.2$ AU, and
                         $\min$(MOID$_{Jupiter}$) $\ge 2$ AU.
\end{itemize}
Thus, even if some of them are affected by mean motion resonances it
appears that many Mars crossing NEOs remain far from close encounters
with, and are not affected by perturbations from, the Earth and
Jupiter. A close approach with Mars must be very deep to cause a
significant perturbation.

Thus, there are islands of long-term stability in the NEO orbital
element phase space that might `collect' NEOs in a manner that could
mimic the appearance of a NEO family.  The \citet{bot2002a} NEO model
can not accurately model these regions of stability because of its low
resolution in the orbital element phase space, the relatively small
number of objects that were originally used to map out the NEOs'
residence time probability distribution, and because it is restricted
to just 3 of the 6 orbital elements.

\section{Conclusions}
\label{sec:conclusions}

We searched for genetic asteroid families in the NEO population using the method proposed by \citet{fu2005} based on identifying clusters of objects with similar orbits.  We enhanced the method's utility by developing a technique for assessing the statistical significance of the identified clusters using 1,000 realistic family-free fuzzy-real NEO orbital element models.  We created our NEO models by cloning members of the known NEO population in a natural way based on the `distance' between each member and its nearest neighbor.  The technique identified three clusters of four or more NEOs among the orbits in the mpcorb.dat data base.  None of the clusters are statistically significant at $\ge3\sigma$ and we conclude that there are as yet no identified families in the NEO population.

The most statistically significant cluster, C1, contains four objects all with $H<20$ and well-determined orbits with the largest member being asteroid 2000~HW$_{23}$.  We performed several additional tests of the C1 cluster's family veracity including checking the members' taxonomic identification, their bias corrected size-frequency distribution, the possibility that they originated in a family-producing tidal disruption at Mars about 71k years in the past, and whether it is also identifiable as a cluster in proper element space.  None of these tests exclude the possibility that C1 is a genetically related family but at the same time none of the tests provide sufficient evidence to elevate the cluster to family status.

The search for families amongst the NEO population is clearly not as straightforward as the same search amongst the main belt population.  Special care must be taken when assessing the statistical significance of a purported NEO family especially with regard to accounting for observational selection effects in the NEO population.  At the very least, mere similarity of the orbital elements as evidenced by the \DSH\ criterion being less than an arbitrary and commonly used value like 0.2 is insufficient for deciding upon any genetic relationship between NEOs and, by extrapolation, between NEOs and meteors.

\newpage

\section*{Acknowledgments}

This work was supported by NASA NEOO grant NNXO8AR22G. MG was also
funded by grants \#136132 and \#137853 from the Academy of Finland.
We acknowledge CSC-IT Center for Science Ltd. for the allocation of
computational resources. Schunova's work was also funded by The
National Scholarship Programme of Slovak Republic for the Support of
Mobility of Students, PhD Students, University Teachers and
Researchers and VEGA grant No. 1/0636/09 from the Ministry of
Education Of Slovak Republic.  We also thank several colleagues who
provided insight into NEO family issues including Kevin Walsh,
Giovanni Valsecchi, Alessandro Morbidelli, Patrick Michel, Joseph
Masiero, Davide Farnocchia, and Seth A. Jacobson.

\newpage

\bibliographystyle{agsm}
\bibliography{references}

\clearpage
\begin{table}
\begin{center}
\begin{tabular*}{\textwidth}{lrrrrrrrrrr}
\hline
Name & $a$ & $\Delta a$   & $e$ & $\Delta e$ & $i$    & $\Delta i$    & $\omega$ & $\Delta \omega$ & $\Omega$ & $\Delta \Omega$\\
     & AU  & $\times10^{-7}$ &     & $\times10^{-6}$  & deg & $\times10^{-5}$ & deg   & $\times10^{-4}$   & deg   & $\times10^{-4}$\\
\hline
\hline
\multicolumn{11}{l}{\textbf{C1}} \\
\hline
2000~HW$_{23}$ & 2.154 & 1 & 0.424 & 1 & 7.76 & 7 & 245 & 2 & 47 & 1\\ 
2006 JU$_{41}$ & 2.123 & 5 & 0.429 & 2 & 7.61 & 8 & 236 & 3 & 52 & 2\\ 
2001 PF$_{14}$ & 2.120 & 4 & 0.410 & 0.5  & 6.78 & 2 & 254 & 3 & 38 & 1\\ 
2008 LN$_{16}$ & 2.141 & 5 & 0.421 & 0.2  & 7.82 & 2 & 261 & 4 & 36 & 2\\
\hline
\hline
\multicolumn{11}{l}{\textbf{C2}} \\
\hline
1999 YD        & 2.463 & 4000 & 0.593 & 70 & 1.38 & 12 & 62 & 0.2 & 10 & 20\\
2003 UW$_{5}$  & 2.470 & 2$\times10^{5}$ & 0.577 & 300   & 1.87 & 77 & 52 & 0.2 & 17 & 10\\ 
2008 UT$_{5}$  & 2.279 & 2$\times10^{5}$ & 0.555 & 400   & 1.61 & 91 & 33 & 0.1 & 38 & 10\\
2008 YW$_{32}$ & 2.319 & 5$\times10^{5}$ & 0.559 & $10^{4}$  & 0.99 & 125 &  2 & 0.6 & 71 & 40\\
2007 YM        & 2.584 & 2$\times10^{7}$ & 0.617 & 3$\times10^{5}$ & 0.99 & 2$\times10^{5}$ & 11 & 0.8 & 60 & 7$\times10^{4}$\\
2005 WM$_{3}$  & 2.674 & 3$\times10^{5}$ & 0.620 & 400 & 1.23 & 56 & 190& 0.3 & 240 & 10\\
\hline
\hline
\multicolumn{11}{l}{\textbf{C3}} \\
\hline
2008 EA$_{9}$   & 1.059 & 342   & 0.080 & 38   & 0.42 & 23   & 336 & 30 & 129 & 30\\
2010 JW$_{34}$  & 0.983 & 53  & 0.055 & 10    & 2.26 & 42 & 43 & 7 & 50 & 10\\
1991 VG         & 1.027 & 1 & 0.049 & 0.2 & 1.45 & 55  & 25 & 0.1 & 74 & 0.9\\
2009 BD         & 1.063 & $10^{7}$    & 0.052 & 48    & 1.27 & 63  & 317 & 10 & 253& 0.06\\
2006 RH$_{120}$ & 1.033 & $10^{7}$   & 0.024 & 21    & 0.60 & 9    & 10 & 600 & 51 & 0.4 \\ 
\hline 
\end{tabular*} 
\end{center}
\caption{: Orbital elements for the members of 3 NEO clusters (C1, C2 \&
  C3) with four or more members.  Members of the clusters are listed
  in order of decreasing size.  None of the clusters are statistically
  significant at $>3\sigma$.  Data is from the JPL Small-Body
  Database Browser {\tt
    http://ssd.jpl.nasa.gov/sbdb.cgi}.} \label{tab:NEOClusterOrbitElements}
\end{table}

\clearpage
\begin{table}
\begin{center}
\begin{tabular*}{0.55\textwidth}{lrrrrr}
\hline
Name & $H$ & $D_S$ & $D_C$ & $D_{SH}$ & $\Delta D_{SH}$ \\
     & mag & m     & m    &           & $\times10^{-6}$  \\
\hline
\hline
\multicolumn{6}{l}{\textbf{C1}} \\
\hline
2000~HW$_{23}$ & 18.5 & 590 & 1,500 & -     & 4\\ 
2006 JU$_{41}$ & 19.3 & 410 & 1,100 & 0.041 & 6\\ 
2001 PF$_{14}$ & 19.5 & 370 & 1,000 & 0.032 & 4\\ 
2008 LN$_{16}$ & 19.9 & 311 &   803 & 0.048 & 4\\
\hline
\hline
\multicolumn{6}{l}{\textbf{C2}} \\
\hline
1999 YD        & 21.1 & 180 & 460 & -     & 80 \\
2003 UW$_{5}$  & 24.3 &  40 & 110 & 0.051 & 3\\ 
2008 UT$_{5}$  & 24.5 &  40 & 100 & 0.044 & 3\\
2008 YW$_{32}$ & 25.2 &  30 &  70 & 0.049 & 10\\
2007 YM        & 26.2 & 180 &  40 & 0.052 & 1600\\
2005 WM$_{3}$  & 27.7 &   8 &  20 & 0.026 & 170\\
\hline
\hline
\multicolumn{6}{l}{\textbf{C3}} \\
\hline
2008 EA$_{9}$   & 27.7 & 8 & 22 & 0.039 & 4\\
2010 JW$_{34}$  & 28.1 & 7 & 18 & 0.050 & 2\\
1991 VG         & 28.5 & 5 & 15 & -     & 270\\
2009 BD         & 28.8 & 5 & 13 & 0.050 & $10^{5}$\\
2006 RH$_{120}$ & 29.5 & 3 &  9 & 0.049 & $10^{5}$\\ 
\hline 
\end{tabular*} 
\end{center}
\caption{: Absolute magnitude ($H$), diameter ($D_S$ and $D_C$) and \DSH\ for the members of 3 NEO clusters (C1, C2 \& C3) with four or more members.  Members of the clusters are listed in order of decreasing size.  None of the clusters are statistically significant at $>3\sigma$.  \DSH\ and its uncertainty were
calculated using the orbital elements and associated errors from {\tt www.jpl.org} as shown in Table~\ref{tab:NEOClusterOrbitElements}.  The diameters $D_S$ and $D_C$ were calculated using albedos of $p_V=0.2$ and
$p_V=0.03$ corresponding to the mean albedos of S and C class asteroids as reported by \cite{mainzer2011}.} 
\label{tab:NEOClusterPhysicalParameters}
\end{table}

\clearpage
\begin{table}[!t]
\begin{center}
\begin{tabular}{c|c|c}
\hline
name & $\min$(MOID$_{Earth}$) & $\min$(MOID$_{Jupiter}$) \\
     &          AU          &         AU           \\
\hline
\hline
2000 HW$_{23}$  & 0.26782  & 2.16112 \\
2001 PF$_{14}$  & 0.27176  & 2.23225 \\
2006 JU$_{41}$  & 0.23183  & 2.18723 \\
2008 LN$_{16}$  & 0.26740  & 2.18886 \\
2002 RA$_{182}$ & 0.27768  & 2.16264 \\
2004 TE$_{18}$  & 0.26278  & 2.23812 \\
\hline
\end{tabular}
\end{center}
\caption{: Minimum MOID with the Earth and Jupiter for the NEOs in the
  C1 and C1$_p$ clusters during the course of their secular evolution.}
\label{tab:minmoid}
\end{table}

%\clearpage
%\listoffigures

\clearpage

\clearpage

\clearpage

\clearpage

\clearpage

\clearpage

\clearpage

\clearpage

\clearpage

\end{document}